\documentclass{article}
\usepackage{setspace}
\usepackage{graphicx} 
\usepackage{natbib}
\usepackage{float}
\usepackage{framed}
\usepackage{natbib}
\usepackage{comment}
\usepackage{hyperref}
\usepackage{subcaption}
\usepackage{geometry}
\usepackage{multirow}
\usepackage{booktabs}
\usepackage{comment}
\usepackage{xcolor}
\usepackage{amssymb}
\usepackage{amsmath}
\usepackage{graphicx}
\usepackage{natbib}
\usepackage{xcolor}
\definecolor{darkgreen}{rgb}{0.0, 0.5, 0.0}
\definecolor{darkorange}{rgb}{0.8, 0.4, 0.0}

\title{\textbf{Optimal Intervention for Self-triggering Spatial  Networks with Application to Urban Crime Analytics}}
\author{Pramit Das, Moulinath Banerjee, Yuekai Sun \\
Department of Statistics, University of Michigan}
\date{\today}

\begin{document}

\maketitle

\begin{abstract} 
\noindent In many network systems, events at one node trigger further activity at other nodes, e.g., social media users reacting to each other's posts or the clustering of criminal activity in urban environments. These systems are typically referred to as \textit{self-exciting networks}. In such systems, targeted intervention at critical nodes can be an effective strategy for mitigating undesirable consequences such as further propagation of criminal activity or the spreading of misinformation on social media.  In our work, we develop an optimal network intervention model to explore how targeted interventions at critical nodes can mitigate cascading effects throughout a Spatiotemporal Hawkes network. Similar models have been studied previously in the literature in \textit{purely temporal} Hawkes networks, but in our work, we extend them to a spatiotemporal setup and demonstrate the efficacy of our methods by comparing the post-intervention reduction in intensity to other heuristic strategies in simulated networks. Subsequently, we use our method on crime data from the LA police department database to find neighborhoods for strategic intervention to demonstrate an application in predictive policing. 
\newline


\end{abstract} 

\section{Introduction to Intervention in Spatiotemporal Hawkes networks}

\subsection{Introduction to Self-exciting processes}

In nature and human activity, we often observe processes where each event can trigger future occurrences, often in nearby regions and within a short timeframe. Examples include aftershocks triggered by a major earthquake \citep{JASAearthquake}, crime activity in one region triggered by crimes in a nearby region \citep{UCLAcrimepaper}, and one neuron firing triggering another neuron to fire \citep{jovanovichawkes2015cumulants}, etc.. Standard non-homogeneous Poisson processes assume events occur independently of past events and are unsuitable for modeling such phenomena. Hawkes processes \citep{hawkes1974cluster}, on the other hand, incorporate the concept of self-excitation (or self-triggering), meaning that the occurrence of an event increases the likelihood of subsequent events. This makes them suitable for modeling phenomena that exhibit clustering and contagion-like behaviors  \citep{Hawkeslikelihoodozaki1979maximum}. In multivariate Hawkes processes, multiple nodes or locations are involved, with events at one node potentially triggering events at other nodes. This is particularly useful in modeling complex systems where different agents interact with each other \citep{uclaSThawkes}. Here, for a network system with $n$ nodes, the intensity at each node $i$ depends not only on its own past events but also on events from other nodes. This is expressed as :

\begin{equation}
\lambda_i(t,x,y|\mathcal{H}_t) = \mu_i(x,y) + \sum_{j=1}^n \sum_{e \in Z^j, t_e < t} \phi_{ij}(t - t_e, x - x_e, y-y_e) \quad \forall i = 1, \dots, n.
\end{equation}

Here, $\lambda_i(t,x,y)$ represents the intensity at node $i$, while $\mathcal{H}_t$ stands for the history of the process till time $t$, $Z^j$ stands for the set of events at node $j$, $\mu_i(x,y)$ is the baseline rate at node $i$ for the location $(x,y)$, and $\phi_{ij}(t - t_e, x - x_e, y-y_e)$ represents the influence of an event from node $j$ at $(t_e,x_e,y_e)$ on triggering an event at $(t,x,y)$ at node $i$. The triggering effect in these models typically diminishes over time and with increasing spatial distance, capturing the intuition that the influence of past events fades as time elapses or as one moves farther away in space.


Applications of these networks span diverse fields such as seismology, environmental science, epidemiology, neuroscience, social media analytics and finance (\citep{JASAearthquake}, \citep{hawkesmalaria}, \citep{HawkesInterventionMehrdad2018discrete}, \citep{hawkes2018hawkesfinance}, \citep{mohlerCOVID2022hawkes}, \citep{Twittermohler2020hawkes}), etc. For instance, in analyzing crime propagation, to study how crimes in one neighborhood trigger other neighborhoods, one could model the system as a multivariate Hawkes network with each neighborhood being a node. We shall consider such a case while analyzing crime data from Los Angeles, California, in Section 4.

\subsection{Intervention in Spatiotemporal Hawkes Networks}

Consider a spatiotemporal Hawkes network with $n$ nodes denoted by, 
$$
N(t,x,y)=\left(N_1(t,x,y), \cdots, N_n(t,x,y)\right)^{\top},
$$ where  $N_i(t,x,y)$ is the counting process corresponding to the events at node $i$ , and the intensity at node $i$ is given by

\begin{equation}
\label{eqn:lambda_network_exp}   
\lambda_i(t,x,y)  =\mu_i(x,y)+\sum_{j=1}^n \int_0^t \phi_{i j}(t-t',x-x',y-y') d N_j(t',x',y').
\end{equation}

Here, the intensity at a specific node $i$ is not only affected by the past events at node $i$, but also by the history of all other nodes. So, by strategically intervening at some of these nodes, we can study the modified behavior of the system and gain valuable insights into how to control or mitigate undesirable events, optimize resource allocation, or improve decision-making in various domains. For example, in epidemiology, intervention strategies can involve targeted vaccination drives or quarantine measures to mitigate the spread and impact of infectious disease outbreaks. In predictive policing, understanding the future impact of interventions in certain neighborhoods on the rest of the city can help us allocate patrol resources effectively. In the context of social media, this could mean restricting the reach of troll accounts to prevent the spread of misinformation.


Suppose the cost of intervention at node $i$ is given by $c_i$. Given the history, i.e., events at each node till time $\tau$, the proposed time of intervention, and a budget $\mathcal{B}$, two natural questions arise:

\begin{itemize}

    \item Which nodes should we intervene at in order to minimize the \textit{expected total rate at time $T$} at all nodes?

    \item  At which nodes should we intervene to minimize the \textit{expected total number of future events till time $T$} over all nodes?
\end{itemize}


 Note that these two problems are related but differ because minimizing the rate at time $T$ focuses on reducing the final total rate across the nodes at time $T$ while minimizing the total number of events up to time $T$ involves supresing average activity throughout the entire event cascade over the time horizon $[\tau, T]$.  For example, consider crime prevention in a city: if we use data from the first week of a month (up to time $\tau=7$ days) to intervene, we might choose locations that reduce overall crime throughout the month for the entire period of $T=30$ days.        (Problem 2). However, if we want to minimize the crime rate at the end of the month (Problem 1), we might target hotspots that are currently active, even if they weren’t major contributors earlier. The key difference is that Problem 1 emphasizes late-stage activity, while Problem 2 prioritizes early triggers that influence the full-time horizon.

\citep{HawkesInterventionMehrdad2018discrete} worked on these problems for purely temporal Hawkes networks. We extend their work to a more general set-up in the setting of spatiotemporal regimes. 
To answer these questions, we obtain analytical expressions for the expected post-intervention intensities and integrate them to obtain the total expected number of events in the system,  post-intervention. Our main contributions are as follows:

\begin{itemize}
    \item We develop a new optimal intervention framework for spatiotemporal Hawkes networks, while previous work by \citep{HawkesInterventionMehrdad2018discrete} considered purely temporal Hawkes networks and did not incorporate spatial information. 

\item We explore more realistic intervention schemes that account for events that continue to have triggering effects after the intervention and diminished post-intervention background intensity at the intervened nodes by introducing quantities such as the \textit{post-intervention survival probability} and the \textit{post-intervention background rate diminishing factor}, respectively, that quantify the effectiveness of the intervention mechanism. Tuning these two parameters can help us explore the full range of interventions with various levels of strength.

\item Due to the triggering nature of crime activity (\citep{UCLAcrimepaper}), the above-mentioned optimization problems have a natural application in predictive policing for patrolling resource allocation under a low to moderate budget scenario. We demonstrate the effectiveness of our algorithm by applying it to real crime data from the Los Angeles Police Department.

\end{itemize}

\section{Methodology}
 For ease of representation, we use vectorized notation. Define the matrix $\Phi(t,x,y)=\left[\phi_{i j}(t,x,y)\right]_{i, j=1 \ldots n}$, vectors $\lambda(t,x,y)=\left(\lambda_1(t,x,y), \ldots, \lambda_n(t,x,y)\right)^{\top}$, and $\mu(x,y)=\left(\mu_1(x,y), \ldots, \mu_n(x,y)\right)^{\top}$. Then, we can compactly rewrite \ref{eqn:lambda_network_exp} in matrix form:
$$
\lambda(t,x,y)=\mu(x,y)+\int_0^t \Phi(t-t',x-x',y-y') d N(t',x',y')
\,.$$

Suppose the background intensity $\mu(x,y)$ integrates to $\mu < \infty$, i.e.
$$
\iint_{\mathbb{R}^2} \mu(x, y)\,dx\,dy = \mu
$$

A commonly used example in the literature (See \citep{NiljanaAkpinarpredpolicing_2021} and \cite{UCLAcrimepaper}) is the Gaussian kernel given by:
$$
\mu_{i}(x,y) = \frac{\mu_i}{2\pi\sigma_0^2}\exp \left\{\frac{-1}{2 \sigma_0^2}\left(x^2+y^2\right)\right\} \forall i \in [n] \,
$$

Note that we are using the above example primarily for illustration, since the spherically symmetric Gaussian spatial kernels are widely used in spatial modeling, denoting a decaying effect as we move away from the center(s). However, the theoretical framework developed in this paper does not depend on this specific format. For example, one could also use a Kernel density estimate involving a mixture of Gaussian densities as the background kernel. 

For the triggering kernel, we adopt a separable form
$$
\begin{aligned}
& \phi_{ij}(t,x,y)=a_{ij}\phi(t,x,y)= a_{ij}\phi_1(t)\phi_2(x,y)\\
& \text{where }  \phi_1(t) =e^{-\omega t} \text{ and } 
\phi_2(x,y)  \text{ is a spatial kernel so that} \iint_{\mathbb{R}^2} \phi_2(x, y)\,dx\,dy = 1.
\end{aligned}.
$$
This splitting leads to decoupling the effect of space and time in the event propagation mechanism. For illustrative purposes, we stick to one of the common kernel choices in the spatiotemporal Hawkes process literature 
$$
\phi_2(x,y)= \frac{1}{2 \pi \sigma^2} \exp \left\{\frac{-1}{2 \sigma^2}\left(x^2+y^2\right)\right\}.
$$ 

Note that the Gaussian form of the spatial kernel captures the intuitive idea that the influence of an event diminishes sharply as we move farther away from its spatial location. Although we present the Gaussian kernel for clarity and its widespread use in applications (\citep{NiljanaAkpinarpredpolicing_2021}, \cite{bernabeu2024_review_hawkesspatio}, \citep{UCLAcrimepaper}, etc.), our theoretical framework is general and accommodates other choices of spatial kernels that satisfy the normalization condition. Several such alternative choices have been proposed in the literature such as:

\begin{enumerate}

\item 
   $$
    \phi(t, x, y) = \frac{K_0}{(t + c)^p} \cdot \frac{e^{\alpha(M - M_0)}}{(x^2 + y^2 + d)^q},
 $$
    where $K_0, p, q, c, d, \alpha$ are parameters. This form captures slower (power-law) decay in time and space, and is well-suited to model phenomena like retaliatory violence or burglary spread that decays non-exponentially. Although originally from seismology, it transfers well to criminology \citep{UCLAcrimepaper} .

    \item 
    \begin{equation}
    \phi_2(x, y) = \frac{1}{2\pi \sigma_x \sigma_y} \exp\left\{-\left(\frac{x^2}{2\sigma_x^2} + \frac{y^2}{2\sigma_y^2}\right)\right\},
    \end{equation}
    where $\sigma_x \neq \sigma_y$ captures directional spread—useful for urban grids or streetscapes where crime spreads more along certain directions.

    \item A coarse-grained kernel given by 
    $$
    \phi(t, x,y) = \frac{1}{(1 + t)(1 +d)}
    $$

   where the distance $d=\sqrt{x^2+y^2}$, as used in \citep{bowers2004prospective}.  Due to its simple mathematical form, it is suitable for real-time policing applications.
  
\end{enumerate}

Apart from parametric formulations, nonparametric methods for describing the intensity of the point process such as kernel density estimates learned via stochastic declustering \citep{UCLAcrimepaper},offer even greater flexibility and allow empirical discovery of intricate spatiotemporal influence patterns. For ease of notation, we subsequently denote $A=((a_{ij}))$. The coefficients $a_{ij}$ determine the strength of influence from node $j$ to node $i$.

\textbf{State of the system: } Define the state of the system at $(t,x,y)$ as 
 $$
 \begin{aligned}
S\left(t,x,y\right)=\frac{1}{2\pi\sigma^2}\int_0^{t} \int_{x' \in \mathbf{R}} \int_{y' \in \mathbf{R}} \exp\left\{-\omega\left(t-t'\right)-\frac{(x-x')^2+(y-y')^2}{2\sigma^2}\right\} d N(t',x',y')
\end{aligned} 
 $$ 
and its temporal version
$$
S(t) = \iint_{(x,y) \in \mathbf{R}^2} S(t,x,y)dxdy \,.$$

And, we denote:
\begin{equation}
\label{eqn:phi_*}
\phi^*(t-t',x-x',y-y') := \frac{1}{2\pi \sigma^2} \exp\left\{-\omega\left(t-t'\right)-\frac{(x-x')^2+(y-y')^2}{2\sigma^2}\right\}.   
\end{equation}

So, in other words, 
$$
 S_i\left(t,x,y\right) = \underbrace{\sum_{\{e \in N^i : t_e < t \}}\phi^*\left(t-t_e, x-x_e, y-y_e\right) }_{\text{contribution of event $e$ in the triggering term}}  \: \: \forall i \in [n]
 $$

In the following study, we consider the sample path till the point of intervention $t=\tau$, i.e., $\mathcal{H}_\tau$, to be given. All expressions henceforth implicitly assume this. In \citep{HawkesInterventionMehrdad2018discrete}'s work, the intervention scheme \textit{deletes} all the events till time $\tau$ at a given subset of nodes $U \subset [n]$ so that they cannot trigger future events any further for $t>\tau$.  But this is usually unrealistic. In reality, intervention mechanisms, such as additional patrolling or increased surveillance, remove only a fraction of events from the affected node. Suppose $(1-p)$ fraction of events where $p \in (0,1)$ is \textit{deleted} from the nodes where intervention is applied. 

We partition the intensity of the process, in the absence of intervention, for $t > \tau$ into two parts:
\begin{equation}
\begin{aligned}
\lambda(t, x, y) &= \mu(x, y) + A S(t, x, y) \\
&= \mu(x, y) + \int_0^t \int_{x' \in \mathbb{R}} \int_{y' \in \mathbb{R}} \frac{A}{2\pi\sigma^2} 
\exp\left\{ -\omega (t - t') - \frac{(x - x')^2 + (y - y')^2}{2\sigma^2} \right\} dN(t', x', y') \\
&= \mu(x, y) + 
\underbrace{\int_0^{\tau} \int_{x' \in \mathbb{R}} \int_{y' \in \mathbb{R}} \frac{A}{2\pi\sigma^2} 
\exp\left\{ -\omega (t - t') - \frac{(x - x')^2 + (y - y')^2}{2\sigma^2} \right\} dN(t', x', y')}_{\text{events before intervention at } \tau} \\
&\quad +
\underbrace{\int_{\tau}^{t} \int_{x' \in \mathbb{R}} \int_{y' \in \mathbb{R}} \frac{A}{2\pi\sigma^2} 
\exp\left\{ -\omega (t - t') - \frac{(x - x')^2 + (y - y')^2}{2\sigma^2} \right\} dN(t', x', y')}_{\text{events after intervention}} \\
&= \mu(x, y) + A e^{-\omega(t - \tau)} S(\tau, x, y) \\
&+ \int_{\tau}^{t} \int_{x' \in \mathbb{R}} \int_{y' \in \mathbb{R}} \frac{A}{2\pi\sigma^2} 
\exp\left\{ -\omega (t - t') - \frac{(x - x')^2 + (y - y')^2}{2\sigma^2} \right\} dN(t', x', y').
\end{aligned}
\end{equation}

\subsection{Behavior in presence of intervention }


Recall the time of intervention is $\tau$ and let the observation period be till time $t=T$.
Denote the whole set of nodes by $V=\{1,2,...,n\}$, $U \subset V$ be the set of nodes intervened at and $\eta_i(t; u)=\mathbb{E}[\lambda_i(t; u)]$, where $t>\tau$ and the expectation is taken over all possible realizations of the process in the time horizon $t \in [\tau,T]$.

Denote an indicator vector $u$ as follows 
to mark which nodes are being intervened on. 

$$
u_i= \begin{cases}0, & i \in U \\ 1, & i \notin U\end{cases}
$$

In the setup described in prior work by \citep{HawkesInterventionMehrdad2018discrete}. it is assumed     that, once a node $i$ is intervened upon, for all events $e \in N_i(\cdot)$ with $t_e < \tau$, the processes corresponding to $\phi_{ki}(t-t_e,x-x_e,y-y_e)$ completely stop $\forall k \in [n].$ In other words, the intervention is assumed to entirely terminate the excitation chains originating from historical events in the nodes that are intervened in. While this makes things analytically simpler, this assumption does not reflect many real-world applications, such as crime prevention or threat mitigation, where intervention efforts may only partially disrupt the influence of prior events. \newline 
To model this more realistically, we assume that after intervention at a node $i \in U$, for each of the events $e \in N_i(\cdot)$ till time $\tau$, the triggered processes would stay active with probability $p$ and would be stopped with probability $(1-p)$. Henceforth, we shall refer to $p$ as the \textit{post-intervention survival probability}. This probabilistic formulation allows for partial retention of triggering power after intervention, capturing the fact that even after proactive measures are taken, the lingering effects of past events till $t=\tau$ may persist. Also, the parameter $p$  provides a flexible way to tune the strength of intervention, more accurately reflecting the incomplete suppression often observed in practical intervention settings.

Now, we mathematically express this scenario after the intervention at time $\tau$. Define $\tilde{S}(t, x, y;u)$ as the $n \times 1$ matrix, whose $i^{\text {th }}$ row is given by

$$
\begin{aligned}
\tilde{S}^i(t, x, y;u) =  
\underbrace{
\sum_{\{e \in N^i : t_e < t \}} \phi^*\left(t-t_e, x-x_e, y-y_e\right) Z_e
}_{\substack{\text{contribution of event $e$ in the triggering term} \\ \text{multiplied by a random binary variable}}}
\end{aligned}
$$

where 
$$
Z_e=1 \text{ if } e \in N^i \text{ and } i \notin U 
$$
and for $e \in N^i \text{ and } i \in U,$
$$
Z_e=\left\{\begin{array}{lll}0 & \text { with probability } & (1-p) \\ 1 & \text { with probability  } & p \end{array}\right. 
$$.

Therefore, for $t>\tau$, after intervention, we can write:
$$
\begin{aligned}
&\lambda(t,x,y ; u) = \mu(x,y)+A e^{-\omega\left(t-\tau\right)} \tilde{S}(\tau, x, y;u) \\
&+\int_\tau^{t} \int_{x' \in \mathbf{R}} \int_{y' \in \mathbf{R}} \frac{A}{2\pi\sigma^2} \exp\left\{-\omega\left(t-t'\right)-\frac{(x-x')^2+(y-y')^2}{2\sigma^2}\right\} d N(t',x',y';u)\,.
\end{aligned}
$$

It is not unnatural to expect that, after being subjected to intervention, such as being given prioritized access to vaccination or utilizing enhanced patrolling resources to prevent crimes, the inherent background rates in the nodes $i\in U$ should decrease, thereby affecting the dynamics of the entire network. However, the described intervention model in \citep{HawkesInterventionMehrdad2018discrete} only considers the reduction in triggering effects post-intervention, but doesn't address the possibility of reduced background rate in the nodes where interventions are performed. We can address this post-intervention background dampening using forms such as $\mu^i(x,y;u) = \gamma\mu^i(x,y)$ with $\gamma\in (0,1)$ 
 or $\mu^i(x,y;u) = \mu \exp\{-\frac{x^2+y^2}{2\sigma_1^2}\}$ where $\sigma_1^2 < \sigma_0^2$ for the nodes $i \in U$. The first one models the post-intervention rate as a fraction of the original background rate, and the second suggests that the `sphere of influence' has decreased post-intervention. 


Here, we consider  the following formulation: 
$$
\tilde{\mu}^i(x, y; u)=\left\{\begin{array}{l}
\gamma \mu^i(x, y), i \in U \\
\\

\mu^i(x, y), i \notin U
\end{array}\right.
$$

for some constant $\gamma \in (0,1)$, which denotes the strength of the dampening effect. Now we shall work out the analytical expressions for the expected total intensity and expected number of events at each node after the intervention.  Using superposition,  we decompose the process $N(t,x,y; u)$ into two independent processes:
$$
N(t,x,y ; u)=N_e(t,x,y ; u)+N_h(t,x,y ; u)
$$
where $N_e(t,x,y; u)$ is the process of events generated from the background intensity $\mu(\cdot)$ from $\tau$ to $t$ and offspring thereof, and $N_h(t,x,y; u)$ denotes the process with the events generated due to the triggering effect of previous events from $\mathcal{H}_\tau$. For notational purposes, define the corresponding temporal processes as $N_e(t;u)$ and $N_h(t;u)$ respectively. We now write down the expressions for the intensities of these processes. 
\begin{equation}
\label{eqn:post_intervention_lambda}
\begin{aligned}
\lambda_h(t,x,y;u) &= \underbrace{A e^{-\omega(t-\tau)}\tilde{S}(\tau,x,y;u)}_{\text{from events before } \tau} \\
&\quad + \underbrace{\int_{\tau}^{t} \int_{x' \in \mathbb{R}} \int_{y' \in \mathbb{R}} \frac{A}{2\pi\sigma^2} \exp\left\{-\omega(t-t') - \frac{(x-x')^2+(y-y')^2}{2\sigma^2}\right\} dN_h(t',x',y';u)}_{\text{from new events generated by history}},
\end{aligned}
\end{equation}

\begin{equation}
\label{eqn:post_intervention_lambda_e}
\begin{aligned}
\lambda_e(t,x,y;u) &= \tilde{\mu}(x,y;u) \\
&\quad + \int_{\tau}^{t} \int_{x' \in \mathbb{R}} \int_{y' \in \mathbb{R}} \frac{A}{2\pi\sigma^2} \exp\left\{-\omega(t-t') - \frac{(x-x')^2+(y-y')^2}{2\sigma^2}\right\} dN_e(t',x',y';u).
\end{aligned}
\end{equation}

where $\tilde{\mu}(\cdot)$ is the post-intervention dampened background intensity as described before. 

Consider the temporal intensities: 

$$
\begin{aligned}
&\lambda_e(t ; u)=\iint_{\mathbb{R}^2} \lambda_e(t, x, y;u)dxdy\\
& \lambda_h(t ; u)=\iint_{\mathbb{R}^2} \lambda_h(t, x, y;u)dxdy    
\end{aligned}
$$
and define the expected temporal expectations post-intervention as:
$$
\eta_e(t;u):=E[\lambda_e(t;u)] = \iint_{\mathbb{R}^2} E\,[\lambda_e(t, x, y;u)]dxdy = \iint_{\mathbb{R}^2} \eta_e(t, x, y ; u) d x d y\,,$$
where $\eta_e(t, x, y; u) := E\,[\lambda_e(t, x, y;u)]$
and 
$$
\eta_h(t;u):=E[\lambda_h(t;u)]=\iint_{\mathbb{R}^2} \eta_h(t, x, y ; u) d x d y \,.
$$
Note that, 
$$
\begin{aligned}
& E\left(d N_e(t, x, y ; u)\right) =E\left[E\left(d N_e(t, x, y) \mid \lambda_e(t, x, y ; u)\right]\right.= E\left[\lambda_e(t, x, y ; u) d t d x d y\right] = \eta_e(t, x, y ; u) d t d x d y\\
& \implies E[dN_e(t;u)]=E\left[\int_{\mathbb{R}^2} d N_e(t, x, y ; u)\right] = \iint_{\mathbb{R}^2} \eta_e(t, x, y ; u) d x d y =\eta_e(t;u).
\end{aligned}
$$

Now, upon integrating out $(x,y)$ from both sides of \ref{eqn:post_intervention_lambda_e}, and taking expectations yields:

\begin{equation}
\label{eqn:eta_e(t,u)}
\eta_e(t ; u)=(\mu \tilde{\textbf{1}})\circ \rho+\int_\tau^t A e^{-\omega\left(t-s\right)} \eta_e(t;u)dt    
\end{equation}

where $\rho$ is a $n \times 1$ vector given by
$$
\rho_i= \gamma \mathbb{I}_{i \in U}+ 1\mathbb{I}_{i \notin U}
$$
In other words, we can write $\rho = \gamma + (1-\gamma)u$ for the intervention vector $u$.

Solving the differential equation \ref{eqn:eta_e(t,u)} using Lemma $1$ from Section $3$ in \citep{farajtabar2016multistage}: 
$$
\begin{aligned}
&\eta_e(t ; u)=\left(I+A(A-\omega I)^{-1}\left(e^{(A-\omega I) t}-I\right)\right) \mu \tilde{1} \circ \rho :=\Psi(t) (\mu \mathbf{1} \circ \rho)\\
&\Longrightarrow \eta_e(t ; u) = \mu \gamma \Psi(t)\tilde{\textbf{1}}+ (1-\gamma) \mu\Psi(t)u    
\end{aligned}
$$
where 
$$
\Psi(t)=I+A(A-\omega I)^{-1}\left(e^{(A-\omega I) t}-I\right).
$$

Similarly, we can get an expression for $\eta_h(t;u)$ given by
$$
\begin{aligned}
\eta_h(t ; u)=\Xi(t-\tau)A(\nu \circ S(\tau))
\end{aligned}
$$
where $\Xi(t)=e^{(A-\omega I) t}$ and $\nu \in \mathbf{R}^{n \times 1}$ is a vector such that $\nu_i = \mathbb{I}_{i \notin U}+p\mathbb{I}_{i \in U} \iff \nu= p+(1-p)u.$

We can interpret $\nu$ as the average likelihood of an event in a node $i \in U$ surviving after intervention.

So, we get 
\begin{equation}
\begin{aligned}
\eta(t; u) &= \eta_e(t; u) + \eta_h(t; u) \\
&= \mu \gamma \Psi(t) \mathbf{1} + (1-\gamma)\mu \Psi(t) u + e^{(A-\omega I)(t-\tau)} A \left( p S(\tau) + (1-p)(u \circ S(\tau)) \right).
\end{aligned}
\label{eqn:eta_expected}
\end{equation}

To compute the total number of expected events till the $T$, one can integrate to get
$$
E\left[N(T ; u)\right]=\int_0^T \eta(s ; u) d s. 
$$

Define the anti-derivatives as
$$
\begin{aligned}
& \Upsilon(t)=\int_0^t \Xi(s) d s \Longrightarrow \Upsilon(t)=(A-\omega I)^{-1}\left(e^{(A-\omega I) t}-I\right) \\
& \Gamma(t)=\int_0^t \Psi(s) d s \Longrightarrow \Gamma(t)=I t+A(A-\omega I)^{-1}(\Upsilon(t)-I t).
\end{aligned}
$$
In terms of the above quantities, for any $t> \tau$, we have 
\begin{equation}
\label{eqn:EN_expected}
\begin{aligned}
E[N(t;u)] &= \Gamma(t)(\mu \circ \phi) + \Upsilon(t-\tau) A (\nu \circ S(\tau)) \\
&= \Gamma(t)\left(\gamma \mu + (1-\gamma)\mu u\right) + \Upsilon(t-\tau)A\left(pS(\tau) + (1-p)(u \circ S(\tau))\right) \\
&= \gamma \Gamma(t)\mu + (1-\gamma)\Gamma(t)(\mu u) + p\Upsilon(t-\tau)A S(\tau) + (1-p)\Upsilon(t-\tau)A(u \circ S(\tau)).
\end{aligned}
\end{equation}
\bigskip

\textbf{Effect of $(p,\gamma)$ on the post-intervention network dynamics:} The analytical expressions derived for the post-intervention expected intensity and the expected number of events  (Equations~\ref{eqn:eta_expected} and~\ref{eqn:EN_expected}) offer insight into how intervention parameters influence the system's evolution. Specifically, a lower value of $\gamma$ reflects a stronger dampening of the background intensity at intervened nodes. As a result, both the expected post-intervention event rate and cumulative event count decrease, reflecting a more substantial long-term suppression of spontaneous activity. On the other hand, higher values of the survival probability $p$ imply that a greater proportion of pre-intervention events continue to exert a triggering influence after intervention. This leads to increased post-intervention expected rates and total event counts, as more of the cascade potential remains intact. These parameters can be tuned as needed to control the intervention strength for specific applications.

The influence of $p$ and $\gamma$ on $\eta(t;u)$ and $E[N(t;u)]$ can be quantified through partial derivative analysis too:
\begin{equation}
\label{eqn:partial_derivatives_eta}
\begin{aligned}
\frac{\partial \eta(t ; u)}{\partial \gamma} &= \mu \Psi(t) (\mathbf{1}-u) \geq 0, \
\frac{\partial \eta(t ; u)}{\partial p}&=e^{(A-\omega I)(t-\tau)} A \left((\mathbf{1}-u) \circ S(\tau)\right) \geq 0.
\end{aligned}
\end{equation}
\begin{equation}
\label{eqn:partial_derivatives_EN}
\begin{aligned}
\frac{\partial \mathbb{E}[N(t; u)]}{\partial \gamma} &= \mu \Gamma(t) (\mathbf{1}-u) \geq 0, \
\frac{\partial \mathbb{E}[N(t; u)]}{\partial p} &= \Upsilon(T-\tau) A \left((\mathbf{1}-u) \circ S(\tau)\right) \geq 0.
\end{aligned}
\end{equation}
The non-negative values of these partial derivatives confirm that reducing $\gamma$ (stronger background suppression) and lowering $p$  (weaker survival of endogenous triggers) both contribute to lowering the future intensity and event count, thereby enhancing the effectiveness of interventions.

\bigskip
\textbf{Spillover effects of Intervention in Hawkes Networks:}
As described in \citep{mohler2015predictive_spillover_intervention}, in some cases, suppressing crime activity in one area leads to reduced crime in other areas as well. This behaviour is naturally captured in our proposed intervention scheme for the Hawkes network model. Specifically, when an intervention is applied at node $i$ at time $\tau$, we downweight both the background intensity by a factor of $\gamma \leq 1$ and the triggering kernel terms associated with past events at node $i$ till time $\tau$ by a multiplicative dampening factor $p \in (0,1)$. After intervention, for $t>\tau$, as seen from \ref{eqn:partial_derivatives_eta} and \ref{eqn:partial_derivatives_EN}, this dampening in turn leads to a reduction in the future expected intensity at time $t$ (and expected number of events till time $t$), not just at node $i$, but also at all other nodes $j \neq i$ that are coupled through the influence matrix $A = (a_{ij})$. So, our intervention mechanism directly accounts for the phenomenon of spillover suppression discussed in \cite{mohler2015predictive_spillover_intervention}.
\bigskip

\textbf{Comments on non-standard Spatial Kernels and Temporal Triggering Functions:}

Note that since the spatial coordinates \((x, y)\) are integrated out from the expressions in equation~\ref{eqn:post_intervention_lambda}, the formulas for the post-intervention expected intensity \(\eta(t;u) = \mathbb{E}[\lambda(t;u)]\) and the expected cumulative number of events \(\mathbb{E}[N(t;u)]\) derived in equations~\ref{eqn:eta_expected} and~\ref{eqn:EN_expected} respectively, remain valid for any arbitrary integrable background rate function \(\mu(x,y)\) and spatial triggering kernel \(\phi_2(x,y)\) satisfying $\iint_{\mathbb{R}^2} \phi_2(x, y) \, dx\, dy = 1.$

This implies that the derived expressions are applicable even when non-Gaussian spatial kernels are used. Such generality is particularly valuable in real-world applications where the background intensity may exhibit irregular structure, for example, due to non-symmetric spatial distributions or geographic heterogeneity. Furthermore, this framework enables flexible modeling with multimodal spatial kernels or non-parametric kernel density estimates, thereby enhancing the model's adaptability to complex spatial patterns observed in practice.

Regarding the temporal triggering function, we adopt an exponential form \(\phi^{ij}_1(t) = a_{ij}e^{-\omega t}\), as it is one of the most widely used choices in the Hawkes process literature (See \citep{Originalhawkes1971spectra}, \citep{Hawkeslikelihoodozaki1979maximum}, \citep{bacry2015hawkes}, \citep{HawkesInterventionMehrdad2018discrete}, etc.). This choice facilitates the derivation of closed-form analytical expressions, such as solving the integral system in equation~\ref{eqn:eta_e(t,u)}. For other possible forms of temporal triggering kernels, while closed-form solutions may not exist, it is possible to numerically compute the corresponding expected intensities, such as \(\eta_e\) and \(\eta_h\), etc., using standard numerical techniques for solving such integral equations.

\subsection{Optimization problems for intervention under resource constraints}

Usually, intervention at any node incurs a non-zero cost; for example, in predictive policing, intervention costs may correspond to the deployment of additional patrolling officers or the expansion of surveillance efforts. As previously discussed, a central challenge in such settings is to determine where and how to intervene in order to suppress undesirable activity while effectively managing limited resources. \citep{HawkesInterventionMehrdad2018discrete} addressed this problem for purely temporal Hawkes processes, without accounting for the spatial locations of events.

Motivated by applications such as algorithmic threat detection and predictive policing—where spatial information plays a critical role—we extend these ideas to spatiotemporal Hawkes networks. Specifically, we formulate two optimization problems aimed at identifying optimal intervention strategies under budget constraints: one focused on minimizing the expected total event rate at a future time point $T$ , and the other on minimizing the expected cumulative number of events up to $t=T$.
Define the quantities 

$$
\begin{aligned}
&\eta_{total}(T;u):= \sum_{i=1}^{n}\eta_i(T; u) \\
&E[N_{total}(T;u)] := \sum_{i=1}^{n}\mathbb{E}\left[N_i(T;u)\right]
\end{aligned}
$$

\begin{enumerate}

\item \textbf{Global rate minimization: \:}Given the intervention time $\tau$ and a pre-decided $T > \tau$, intervention costs $c_i >0  \: \: \forall i \in [n]$ and a total budget $\mathcal{B}$, solve for:

$$
\begin{aligned}
&\min_{U}\eta_{total}(T; u)\\
& \text{ subject to  } \sum_{i=1}^{n} (1-u_i)c_i \leq \mathcal{B}\\
\vspace{0.2cm}
& \text{ and } u \in \{0,1\}^n.    
\end{aligned}
$$

\item \textbf{Global invasion minimization: \:} 

Given the same inputs, solve for

$$
\begin{aligned}
&\min_{U} E[N_{total}(T;u)]\\
& \text{ subject to  } \sum_{i=1}^{n} (1-u_i)c_i \leq \mathcal{B} \\
\vspace{0.2cm}
& \text{ and } u \in \{0,1\}^n.
\end{aligned}
$$
    
\end{enumerate}

As seen from \ref{eqn:eta_expected} and \ref{eqn:EN_expected}, the above-mentioned optimization problems to minimize the expected total rate at time $T$ or the expected total number of events till time $T$ boil down to solving binary programs under budget constraints. In practice, we enforce LP relaxation for the variables as $0 \leq u_i \leq 1 \text{\: for all} \: \:  i =1(1)n$ and formulate them as mixed linear integer programming (MILP) problems to be solved using standard solvers such as CPLEX, Gurobi, SciPy module in Python, etc.

\section{Comparison of Optimal LP-based and Heuristic Intervention Strategies}

 We now numerically illustrate the impact of post-intervention dampening of background intensities on the overall network dynamics through simulations on synthetic networks and a real-world case study using the Los Angeles Police Department crime dataset.

To contrast different intervention strategies, we consider two scenarios: one where post-intervention dampening of background intensities is accounted for and another where it is not. In the latter case, we set $\gamma=1$, meaning only the triggering effect is reduced while the background intensity remains unchanged. By comparing these cases, we aim to assess the influence of background intensity dampening on event suppression and overall network behavior.
\newline
We aim to evaluate and compare different intervention strategies for minimizing both the total rate of events (measured by $\eta_{total}(T;u)$) and the total expected number of events $E[N_{total}(T; u)]$ in a multivariate Hawkes process. Two types of strategies are considered: 
\begin{itemize}
    \item LP-based strategies, which optimize the intervention based on minimizing either the total rate ($\eta$) or the total expected number of events ($E[N(T; u)]$) using linear programming and,
    \item heuristic strategies, where nodes are selected for intervention based on exogenous intensity $\mu_i$ or the total number of events at each node up until the observation time $\tau$.
\end{itemize}

For simulating a multivariate Spatiotemporal Hawkes network, we used $n=200$ nodes, and the spatial dimension is $d=2$. The observation time horizon was set to $\tau =10.$, and the full-time horizon was $T = 2\times \tau$. The background intensity $\mu_i$ for each node was drawn from a uniform distribution $\mathcal{U}(0.01, 0.0.5)$. The interaction matrix $A$ was randomly generated such that each entry $A_{ij} \sim 1.5 \times \mathcal{U}(0, 1)$, representing the triggering effects between nodes. The decay rate is $\omega = 0.2$, and the spatial variances are $\sigma_0^2 = 0.25^2, \sigma^2 = 0.1^2$. To maintain stationarity, if needed, we scale $A$ to ensure the spectral radius $\rho(A) < \omega$. The intervention cost for each node was taken as a constant value of $c=1$, augmented by the total number of events observed at that node up to time $\tau$: 
$$
c_i = c+ N_{i}(\tau)
$$

We evaluate interventions under different parameter settings for \textit{post-intervention survival probability} $p$ and the \textit{post-intervention background intensity scaling factor} $\gamma$. Specifically, we consider the following $6$ combinations: 
\[
(p, \gamma) \in \{0.1,0.3\}\times \{0.6,0.8,1\}
\]
These settings allow us to examine the effect of varying the survival probability of post-intervention events ($p$) and the extent to which the background intensity diminishes post-intervention ($\gamma$). 

The case where $\gamma = 1, p=0 $ corresponds to a setup analogous to \cite{HawkesInterventionMehrdad2018discrete} where background intensities remain unchanged after intervention and all the past events till time $\tau$ in the nodes intervened at stop further triggering activity.

As in \citep{HawkesInterventionMehrdad2018discrete}, we set the total budget to be a fraction of the total cost to intervene in all the nodes. Here,
$$
\mathcal{B} = \left(\sum_{i=1}^n c_i \right)\times \frac{q}{100}.
$$

\begin{figure}[h]
    \centering
    \includegraphics[width=\textwidth]{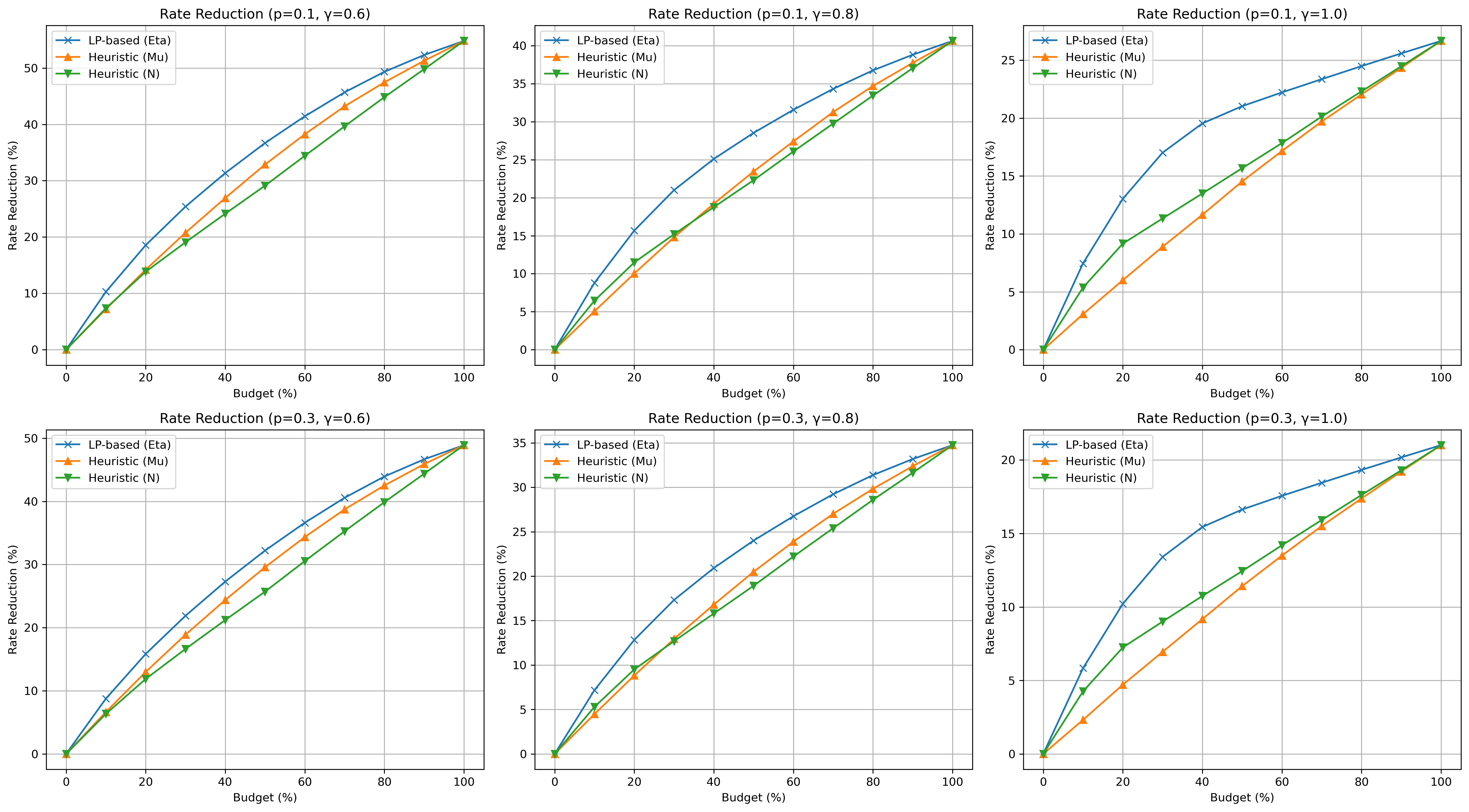}
    \caption{Effect of varying $p$ and $\gamma$ on expected total rate reduction across different budget levels.}
    \label{fig:rate_reduction_results}
\end{figure}

\begin{figure}[h]
    \centering
    \includegraphics[width=\textwidth]{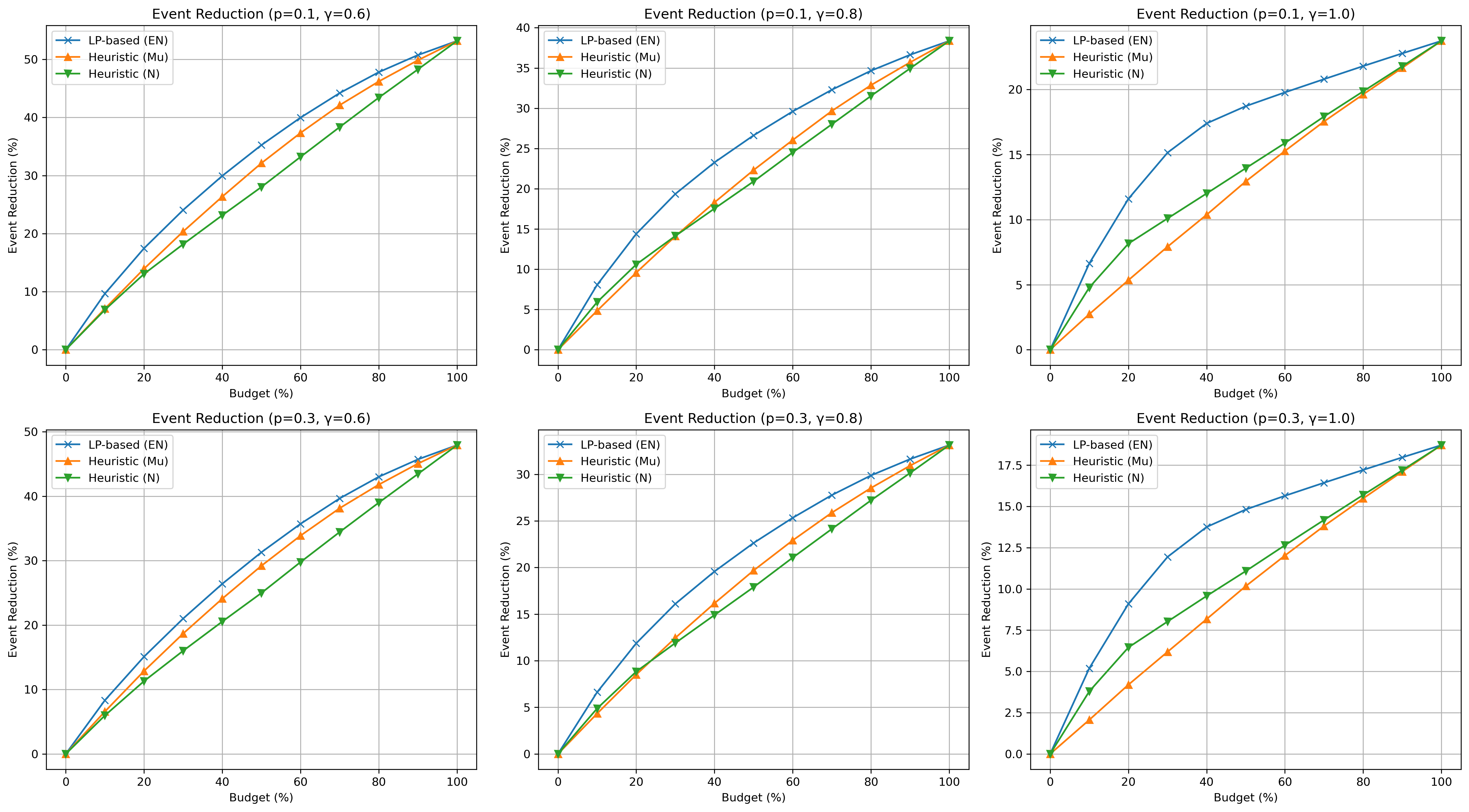}
    \caption{Effect of varying $p$ and $\gamma$ on expected number of events reduction across different budget levels.}
    \label{fig:event_reduction_results}
\end{figure}

For each value of $q \in \{10,20,.., 80, 90 \}$, we ran $n=100$ independent realizations of the multivariate Hawkes process. In each realization, we computed the intervention vector $u$ according to the following strategies \footnote{Note that we colorcode the strategies to use the same color labeling in Figures \ref{fig:rate_reduction_results} and \ref{fig:event_reduction_results}.}:

\begin{enumerate}

    \item (\textcolor{red}{Optimal $\eta_{total}$-based strategy}) The proposed optimal LP-based strategy to minimize the total rate ($\eta_{total}$) after the intervention,
    
    \item  (\textcolor{blue}{Optimal $E(N_{total})$-based strategy}) The proposed optimal LP-based strategy to minimize the total expected number of events ($E[N_{total}(T; u)]$),
    \item  (\textcolor{darkorange}{$\mu_i$-based heuristic strategy}) A heuristic strategy based on exogenous intensity ($\mu_i$) \footnote{Here, $\mu_i=\iint_{\mathbb{R}^2} \mu_i(x, y) d x d y$, the $i^{th}$ entry of the vector $\mu$}, where nodes with the highest $\mu_i$ are prioritized for intervention and
    \item (\textcolor{darkgreen}{$N_i(\tau)$-based heuristic strategy}) A heuristic strategy based on the total number of events ($N_i(\tau)$), where nodes with the highest number of events observed until time $\tau$ are selected for intervention.

\end{enumerate}

For each strategy, two key performance metrics were measured:

\bigskip

\noindent\textbf{1. Percentage reduction in total event rate:}
\[
\% \text{ reduction in expected total rate at $t=T$} 
= \left(1 - \frac{\eta_{\text{total}}(T; u)}{\eta_{\text{total}}(T; u_{\text{no intervention}})}\right) \times 100
\]

\bigskip

\noindent\textbf{2. Percentage reduction in expected total number of events:}
\[
\% \text{ reduction in total number of events till time $T$}
= \left(1 - \frac{\mathbb{E}[N_{\text{total}}(T; u)]}{\mathbb{E}[N_{\text{total}}(T; u_{\text{no intervention}})]}\right) \times 100
\]

Here $u_{\text{no intervention}} = \mathbf{1}$ denotes the vector of all ones (i.e., no intervention at any node).

For each combination of $(p,\gamma)$, the results were averaged over $100$ realizations and were compared across different budget percentages, and two plots \ref{fig:rate_reduction_results}
and \ref{fig:event_reduction_results} were produced.

From the results, we observe that as $\gamma$ decreases (e.g., from $\gamma=1$ to $\gamma=0.8$ or $\gamma=0.6$), there is a higher percentage reduction in both the rate ($\eta$) and expected total number of events $E[N_{total}(T; u)]$. This effect is more pronounced when the budget percentage is high, indicating that the intervention strategies become significantly more effective in suppressing future events when they also reduce the background intensity. The optimal LP-based strategies consistently achieve a greater reduction in expected total rate ( or total number of events) post-intervention compared to heuristic strategies, but all methods benefit from lower $\gamma$, as seen in the sharper decline in event and rate reduction curves.

Varying $p$ allows us to explore different levels of event survival post-intervention, where a lower $p$ corresponds to more effective intervention in reducing immediate triggering effects, while a higher $p$ models a scenario where interventions have a more limited impact on mitigating or suppressing future events. Comparing $p=0.1$ and $p=0.3$ for $\gamma=1$, we see expectedly that a higher value of $p$ leads to a lower reduction in event rates and counts across all budget levels. 

However, the difference is more pronounced for lower budgets, suggesting that when interventions are limited, a higher $p$ makes it harder to control the cascading effects of the Hawkes process. The combination of low $p$ and low $\gamma$ (e.g., $p=0.1, \gamma=0.6$) leads to the most significant reduction in events and rates. The curves in these cases show a steep decline in mitigating further activity, indicating that interventions are highly effective when they not only disrupt triggering dynamics (low $p$) but also dampen the background intensity (low $\gamma$). Conversely, the case $p=0.3, \gamma=1$ exhibits the least reduction in event rates and counts, highlighting that without background suppression and with persistent triggering effects, interventions have a much weaker long-term impact.

Overall, these findings suggest that for optimal intervention strategies, both immediate triggering reduction ($p$) and long-term background suppression ($\gamma$) play crucial roles. The results highlight the importance of designing interventions that not only disrupt immediate event propagation but also lead to sustained reductions in background intensity to achieve maximal suppression of future events. The plots show that, for both  $\gamma=1$ and $0<\gamma<1$, the optimal strategies consistently outperform the heuristic approaches based on intervention at nodes with the highest base rates ($\mu_i$'s) or number of events $N_i(\tau)$, particularly when the budget is between 30\% and 70\% of the total intervention cost at all the nodes. The gap between the LP-based strategies and the heuristic interventions widens significantly in this budget range, indicating the superior efficiency of LP-based optimization in moderately resource-constrained environments. However, as the budget increases to 80\% and above of the total cost required to intervene at all nodes, the performance gap starts to narrow, with the heuristic strategies approaching the performance of the optimal LP-based strategies. This is expected, as at higher budget levels where interventions can be made at almost all nodes, any common-sense strategy tends to perform similarly since there are enough resources to intervene at nearly every critical location.

\section{Case study of crime data from Los Angeles, CA}

Los Angeles is one of the largest cities in the United States and experiences a high number of crimes. Since 2020, the police department has been compiling crime records publicly in a relational database. Each record typically includes details such as the type of crime, the date, time, location, victim information, and various additional descriptors, such as the weapon used or the premises.  We refer to the 
\href{https://data.lacity.org/Public-Safety/Crime-Data-from-2020-to-Present/2nrs-mtv8/data_preview}{LA City Crime Database} for the details. 
\newline
Self-triggering patterns are often observed in various types of crime data, such as burglary and gang violence \citep{UCLAcrimepaper}. Self-exciting point processes, particularly Hawkes models, have been used widely in the literature to model crime data and predict future crime hotspots (\citep{NiljanaAkpinarpredpolicing_2021}, \citep{mohler2014marked}, \citep{mohlerhawkes}, \citep{mohler2019reducingJQC}). Multivariate Hawkes processes have also been used in the criminology literature to understand \textit{spillover effects} (\citep{YaoXiecrime911calltextAtlanta2020spatial}, \citep{leverso2025measuringJQC}). With the rise of machine learning and AI tools, police departments increasingly deploy these techniques to complement traditional policing methods \citep{YaoXiecrime911calltextAtlanta2020spatial}.

\begin{figure}[h]
    \centering
    \includegraphics[width=\textwidth]{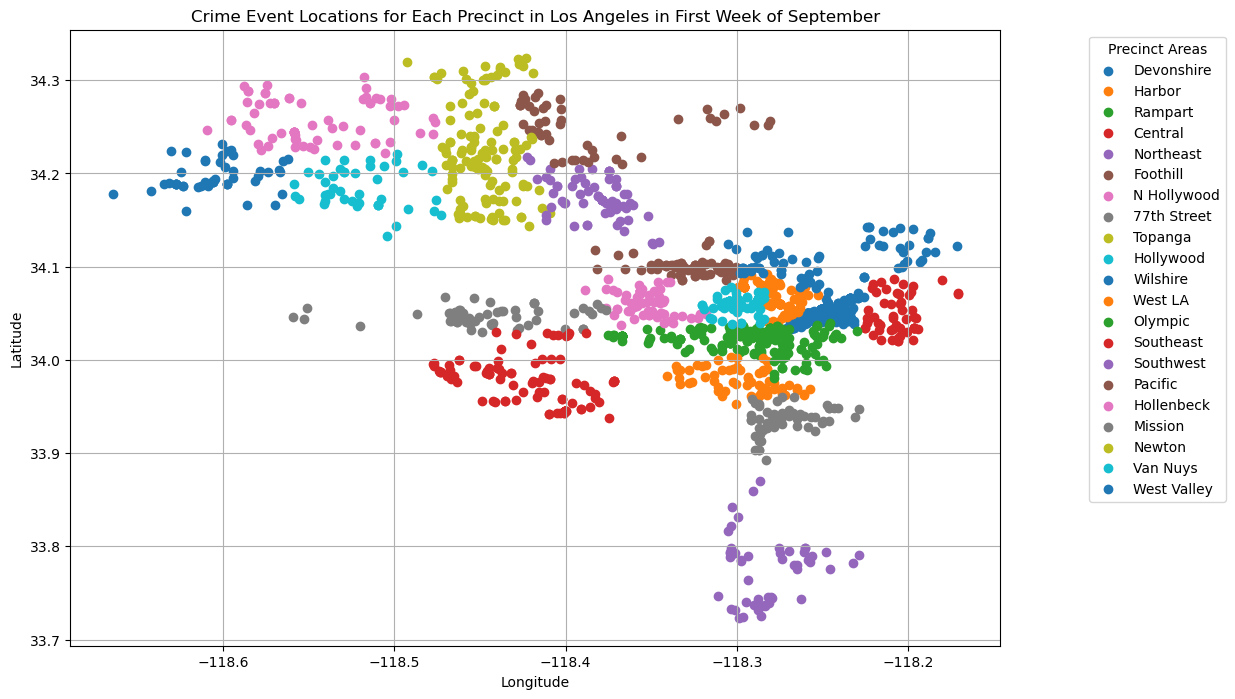}
    \caption{Crime Event Locations for Each Precinct in Los Angeles in the first week of September.}
    \label{fig:LA_crime_locations}
\end{figure}

To fit a spatiotemporal Hawkes process to Los Angeles crime data, we use the model described by  \citep{UCLAcrimepaper}. The entire geographical area of LA is divided into $21$ areas/precincts, and we shall consider these as nodes for our Spatial network model. We downloaded the crime records for the first week of September 2024 \footnote{ The time horizon $[0,\tau]$ is intentionally chosen to be not too large, as crime patterns (and even reporting mechanisms) can change over time, potentially violating the stationarity assumptions of the model.}, which contain $1579$ entries (see Figure~\ref{fig:LA_crime_locations}).

 To model this data as a multivariate spatiotemporal Hawkes network, we adopt the framework described in \cite{uclaSThawkes}.
 Suppose the history of the process till time $\tau$ (point of intervention) is given by 
$\left\{\left(t_i, x_i, y_i, u_i\right)\right\}_{i=1}^N$ \footnote{Here, $u_i \in \{1,2,....,n\}$ denotes the node where the event $(t_i,x_i,y_i)$ occurred.} . As per \cite{uclaSThawkes}, the background intensity is modeled as a Gaussian kernel density estimate as:

\[
\mu^*_u(x, y) = \sum_{i=1}^{N} \frac{\beta_ {uu_i}}{2 \pi \delta^2 \tau} \times \exp\left( -\frac{(x - x_i)^2 + (y - y_i)^2}{2 \delta^2} \right).
\]

Here, \( \delta \) represents the bandwidth parameter controlling the spatial spread of the background kernel, \(\tau\) denotes the total observation time window, and \( \beta_{u u_i} \) quantifies the contribution of an event at node \(u_i\) to the background intensity at node \(u\). Specifically, \( \beta_{u u_i} \) can capture heterogeneity across nodes by weighting the influence of events differently depending on their originating node and the target node. This formulation effectively treats the historical events as a weighted kernel density estimate (KDE), enabling flexible and spatially smooth modeling of the background event intensity in a node-specific manner.

The triggering kernel has the standard product form, featuring an exponential temporal decay and a spatial Gaussian kernel.
\[
\phi(x, y, t) = \phi_1(t) \times \phi_2(x, y) = \omega \exp(-\omega t) \times \frac{1}{2\pi \sigma^2} \exp\left( -\frac{x^2 + y^2}{2\sigma^2} \right),
\]


The intensity function at node $u$ (conditioned on the history of the process) 
is given by:

\[
\lambda_u(t, x, y)=\mu^*_u(x, y)+\sum_{t>t_i} A_{u_i u} \phi\left(x-x_i, y-y_i, t-t_i\right)
\]

The interactions between the events at node $u_j$ with these of node $u_i$ is measured by $A_{u_i u_j}$, where $u_i, u_j \in \{1,2,\cdots, n\}$. We employ an EM algorithm, as described in Section 3 of \citep{uclaSThawkes}, to fit this process to our data by estimating the parameters  $\omega, \sigma, \delta$, $A_{u_iu_j}$'s and $\beta_{uu_i}$'s. In our case, the number of nodes is $n=21$ and the number of events is $N=1579$.

In the context of predictive policing, intervention at a node (i.e., neighborhood) involves strategically enhancing patrols in these areas to deter criminal activities. However, budget constraints are often a consideration due to the substantial resources required. In this section, we demonstrate the application of our proposed method to illustrate how we can select locations for strategic intervention under low to moderate-budget scenarios by using crime data from the city of Los Angeles. The table below compares the different intervention strategies for reducing the total rate 
$$
\sum_{i=1}^{n} \eta(T;u)
$$
compared to the no-intervention scenario subject to different budget constraints given by:

$$
\mathcal{B} = \frac{q}{100} \times \sum_{i=1}^n c_i.
$$

Here again, the intervention cost at node $i$ is assumed to be a base cost plus the number of events in node $i$. This is a natural choice since we can assume that it takes a base level of human resources or money to intervene at a node, and then it increases proportionally with each incident at that node till time $\tau$. Here, the events of the first week of September are considered in the training set using the time scale $\tau$ being one week, and the total horizon $T = 4.3 \tau$ denotes the full month of September. 
\newline
Now, we analyze the performance of our proposed LP-based optimal intervention techniques in the context of crime reduction using LA data. Two cases are considered: (i) $\gamma=1$, where the background intensities remain unaffected post-intervention, similar to the setup studied in the literature by \citep{HawkesInterventionMehrdad2018discrete}; and (ii) $\gamma<1$, where the background intensities are dampened after intervention, reflecting real-world scenarios where successful interventions are expected to reduce future crime risk. Previous work by \citep{HawkesInterventionMehrdad2018discrete} has assumed $p=0$, but this is not realistic. Therefore, for both situations, we set $p$ to be a small positive number, specifically $p=0.1$, to compare the optimal strategies with heuristic ones. \newline
Under the setting $\gamma = 1$, where the background intensities remain unchanged post-intervention, the optimal LP-based strategy demonstrates improvements over heuristic approaches across a range of budget levels. In low-budget scenarios ($q = 10\%$--$20\%$), it achieves a 10--20\% reduction in both the total expected crime intensity and the number of events, outperforming heuristics by approximately 4 percentage points. In moderate-budget regimes ($q = 40\%$--$60\%$), the optimal strategy continues to deliver strong performance, with 39--56\% reductions in crime rate and expected event counts, exceeding heuristic results by 3.5--5\%. Furthermore, the regions of \textbf{Central, Hollenbeck, Harbor, Hollywood, Wilshire, and West Valley} are consistently identified as optimal intervention sites, indicating their central role in influencing downstream crime dynamics in this scenario.
For details, refer to Table \ref{tab:LA_rate_reduction_comparison} and \ref{tab:LA_event_reduction}. 

\begin{table}[h]
    \centering
    \begin{tabular}{|c|c|c|c|}
        \hline
        \textbf{Budget $q$  (\%)} & \textbf{Optimal strategy} & \textbf{$\mu$ based heuristic} & \textbf{$N_i(\tau)$ based heuristic} \\
        \hline
        10.0 & 10.40 & 6.36 & 6.36 \\   
        20.0 & 20.22 & 16.83 & 16.83 \\  
        30.0 & 30.11 & 26.49 & 26.49 \\  
        40.0 & 39.12 & 32.90 & 35.62 \\  
        50.0 & 47.83 & 42.22 & 43.05 \\  
        60.0 & 55.84 & 52.10 & 50.21 \\  
        70.0 & 64.00 & 58.36 & 57.38 \\  
        80.0 & 71.63 & 67.75 & 67.45 \\  
        90.0 & 78.21 & 76.31 & 74.07 \\
        \hline
    \end{tabular}
    \caption{Comparative total expected rate reduction (\%) for optimal vs. heuristic strategies compared to the no intervention scenario using $(p,\gamma)=(0.1,1)$ in the Los Angeles crime data.}
    \label{tab:LA_rate_reduction_comparison}
\end{table}

\begin{table}[h]
    \centering
    \begin{tabular}{|c|c|c|c|}
        \hline
        \textbf{Budget $q$ (\%)} & \textbf{Optimal strategy} & \textbf{$\mu$-based heuristic} & \textbf{$N_i(\tau)$-based heuristic} \\
        \hline
        10.0 & 10.39 & 6.35 & 6.35 \\
        20.0 & 20.21 & 16.82 & 16.82 \\
        30.0 & 30.09 & 26.47 & 26.47 \\
        40.0 & 39.10 & 32.88 & 35.60 \\
        50.0 & 47.79 & 42.19 & 43.03 \\
        60.0 & 55.81 & 52.08 & 50.19 \\
        70.0 & 63.96 & 58.34 & 57.36 \\
        80.0 & 71.59 & 67.72 & 67.42 \\
        90.0 & 78.18 & 76.27 & 74.03 \\
        \hline
    \end{tabular}
    \caption{Reduction in total expected number of events (\%) for different strategies under LA Crime Data for $(p,\gamma)=(0.1,1)$).}
    \label{tab:LA_event_reduction}
\end{table}

However, in the above analysis, we did not incorporate the possibility of diminished background effects after the intervention. Now, we incorporate the diminished post-intervention background effect for the case of $\gamma = 0.75$. In this case, the optimal LP-based strategy retains its advantage, particularly in the low to mid-budget ranges ($q = 10\%$--$20\%$ and $q = 40\%$--$70\%$), where it achieves approximately 2.5--5\% greater reductions in both the total expected rate and number of future events compared to heuristic strategies, as seen in Tables~\ref{tab:LA_rate_reduction_comparison_gamma} and~\ref{tab:LA_num_reduction_comparison_gamma}. In terms of spatial targeting, the optimal intervention sites under this setting include \textbf{Central, Hollenbeck, Harbor, Hollywood, Wilshire, West Valley, Olympic, and Topanga}. Notably, the neighborhoods of Olympic and Topanga did not appear as strategic intervention sites ( in the mid-budget case) in the $\gamma=1$ scenario, illustrating how modeling post-intervention dampening of background rates can influence the prioritization of target locations. Finally, for high-budget scenarios ($q \geq 80\%$), all strategies tend to converge in performance, reflecting diminishing returns when the budget is sufficient to intervene in most of the network.

\begin{table}[h]
\centering
\begin{tabular}{|c|c|c|c|}
\hline
\textbf{Budget ($q$\%)} & \textbf{Optimal LP-based} & \textbf{Heuristic $\mu_i$ based} & \textbf{Heuristic $N_i(\tau)$} \\ \hline
10  & 10.56  & 6.40   & 6.40   \\ 
20  & 20.44  & 16.94  & 16.94  \\ 
30  & 30.43  & 26.69  & 26.69  \\ 
40  & 39.58  & 33.18  & 35.92  \\ 
50  & 48.33  & 42.61  & 43.41  \\ 
60  & 56.48  & 52.61  & 50.66  \\ 
70  & 64.72  & 58.96  & 57.90  \\ 
80  & 72.39  & 68.48  & 68.10  \\ 
90  & 79.08  & 77.14  & 74.80  \\ \hline
\end{tabular}
\caption{Rate Reduction (\%) for Different Strategies compared to the No intervention scenario in LA crime data with $(p,\gamma)=(0.1,0.75)$}
\label{tab:LA_rate_reduction_comparison_gamma}
\end{table}

\begin{table}[h]
\centering
\begin{tabular}{|c|c|c|c|}
\hline
\textbf{Budget (\%)} & \textbf{Optimal LP-based} & \textbf{Heuristic $\mu_i$ based} & \textbf{Heuristic $N_i(\tau)$ based} \\
\hline
10  & 10.55  & 6.39   & 6.39   \\
20  & 20.43  & 16.92  & 16.92  \\
30  & 30.41  & 26.67  & 26.67  \\
40  & 39.56  & 33.17  & 35.90  \\
50  & 48.30  & 42.58  & 43.39  \\
60  & 56.50  & 52.00  & 50.63  \\
70  & 64.69  & 58.94  & 57.89  \\
80  & 72.36  & 68.45  & 68.07  \\
90  & 79.05  & 77.10  & 74.78  \\
\hline
\end{tabular}
\caption{Reduction in expected total number of events (\%)  for different strategies compared to the scenario in the absence of intervention for the LA crime dataset with $(p,\gamma)=(0.1,0.75)$}
\label{tab:LA_num_reduction_comparison_gamma}
\end{table}

Many police departments currently tend to operate under resource constraints (\href{https://www.theiacp.org/sites/default/files/239416_IACP_RecruitmentBR_HR_0.pdf}{See here} for details). This case study illustrates how leveraging data-driven insights for optimal resource allocation can help reduce future crime rates. 

\section{Summary and Discussion}

In this study, we explore optimal intervention strategies in self-exciting spatiotemporal Hawkes networks, where events propagate across nodes through both triggering effects and background activity. Motivated by applications such as predictive policing, threat detection, and epidemiology, we address the challenge of identifying nodes for intervention to minimize future event cascades across the network. Building on prior work focused primarily on purely temporal Hawkes networks \citep{HawkesInterventionMehrdad2018discrete}, we develop a more realistic intervention model by introducing two key quantities: the post-intervention survival probability \(p\) and the background intensity dampening factor \(\gamma\). The parameter \(p\) models the probability that an event continues to trigger future activity even after intervention, thus controlling the persistence of endogenous propagation. The parameter \(\gamma\) captures the reduction in background activity at intervened nodes, reflecting broader deterrence or disruption effects. Together, \(p\) and \(\gamma\) allow us to flexibly model a continuum of intervention effects ranging from partial to near-complete suppression of future activity. Our results demonstrate that LP-based optimization strategies consistently outperform heuristic methods based on exogenous intensities or observed event counts, particularly in low to moderate budget regimes where strategic targeting is critical. Incorporating the effects of post-intervention background dampening (\(\gamma<1\)) leads to greater overall reductions in expected crime rates and event counts, especially when interventions are jointly designed to disrupt both immediate triggering and background-driven processes.

Through simulations and a real-world application to Los Angeles crime data, we illustrate that explicitly accounting for network-based crime propagation dynamics enables smarter and more efficient resource allocation. Our framework shows how predictive policing efforts can be enhanced by embedding network-aware optimization tools into decision-making pipelines. By adjusting intervention parameters \(p\) and \(\gamma\) according to practical considerations or policy goals, agencies can tailor their intervention strategies to balance aggressiveness, fairness, and resource constraints.

\bigskip

For large values of $n$, simulating multiple realizations of multivariate Hawkes networks with $n$ modes is a computationally expensive task. So, in the future, leveraging computing clusters, we plan to extend these experiments for larger network sizes (e.g., $n$ in the order of $ 10^3$ or $ 10^4$, etc.) to assess whether network size influences the relative effectiveness of different intervention strategies. While we work with a fully connected network of nodes here,  many practical settings, not all pairs of nodes in the network interact meaningfully—e.g., crimes in one neighborhood may not affect distant or demographically dissimilar areas. So, setting some $a_{ij}$'s to be $0$ enforces a sparse Hawkes network that reflects domain knowledge such as spatial adjacency or social ties, improves interpretability while reducing the risk of overfitting. Such sparsity can be imposed directly via hard constraints or induced through $\ell_1$-penalization during the estimation \citep{sparsehawkesbacry}. This accelerates computation during both model estimation and downstream tasks. Additionally, incorporating stochastic programming techniques can help model intervention costs $c_i$ or edge strengths $a_{ij} $ as random variables rather than fixed amounts, aiding in building decision-making algorithms under uncertainty. Further research is also needed to integrate fairness considerations in our approaches to inform policies.


\end{document}